\documentclass[manuscript]{acmart}
\AtBeginDocument{%
  \providecommand\BibTeX{{%
    \normalfont B\kern-0.5em{\scshape i\kern-0.25em b}\kern-0.8em\TeX}}}

\setcopyright{none}
\copyrightyear{2024}
\acmYear{2024}

\acmConference[CHI '24 Workshop: With or Without Permission]{}{May 12, 2024}{Honolulu, HI, USA}

\usepackage{xpatch}

\makeatletter
\xpatchcmd{\ps@firstpagestyle}{Manuscript submitted to ACM}{}{\typeout{First patch succeeded}}{\typeout{first patch failed}}
\xpatchcmd{\ps@standardpagestyle}{Manuscript submitted to ACM}{}{\typeout{Second patch succeeded}}{\typeout{Second patch failed}}    \@ACM@manuscriptfalse
\makeatother

\acmBooktitle{Proceedings of With or Without Permission: Site-Specific Augmented Reality for Social Justice Workshop at the CHI Conference on Human Factors in Computing Systems (CHI '24)} 
\acmISBN{}

\usepackage{subfigure}

\begin{document}

\title{ARtivism: AR-Enabled Accessible Public Art and Advocacy}

\author{Lucy Jiang}
\affiliation{%
  \institution{Cornell University}
  \city{Ithaca, NY}
  \country{USA}}
\email{lucjia@cs.cornell.edu}


\begin{abstract}
Activism can take a multitude of forms, including protests, social media campaigns, and even public art. The uniqueness of public art lies in that both the act of creation and the artifacts created can serve as activism. Furthermore, public art is often site-specific and can be created with (e.g., commissioned murals) or without permission (e.g., graffiti art) of the site’s owner. However, the majority of public art is inaccessible to blind and low vision people, excluding them from political and social action. In this position paper, we build on a prior crowdsourced mural description project and describe the design of one potential \textit{process artifact}, ARtivism, for making public art more accessible via augmented reality. We then discuss tensions that may occur at the intersection of public art, activism, and technology.

\end{abstract}

\ccsdesc[500]{Human-centered computing~Accessibility}

\keywords{art, augmented reality, blind, low vision, accessibility, social justice, activism, public art}

\begin{teaserfigure}
  \includegraphics[width=\textwidth]{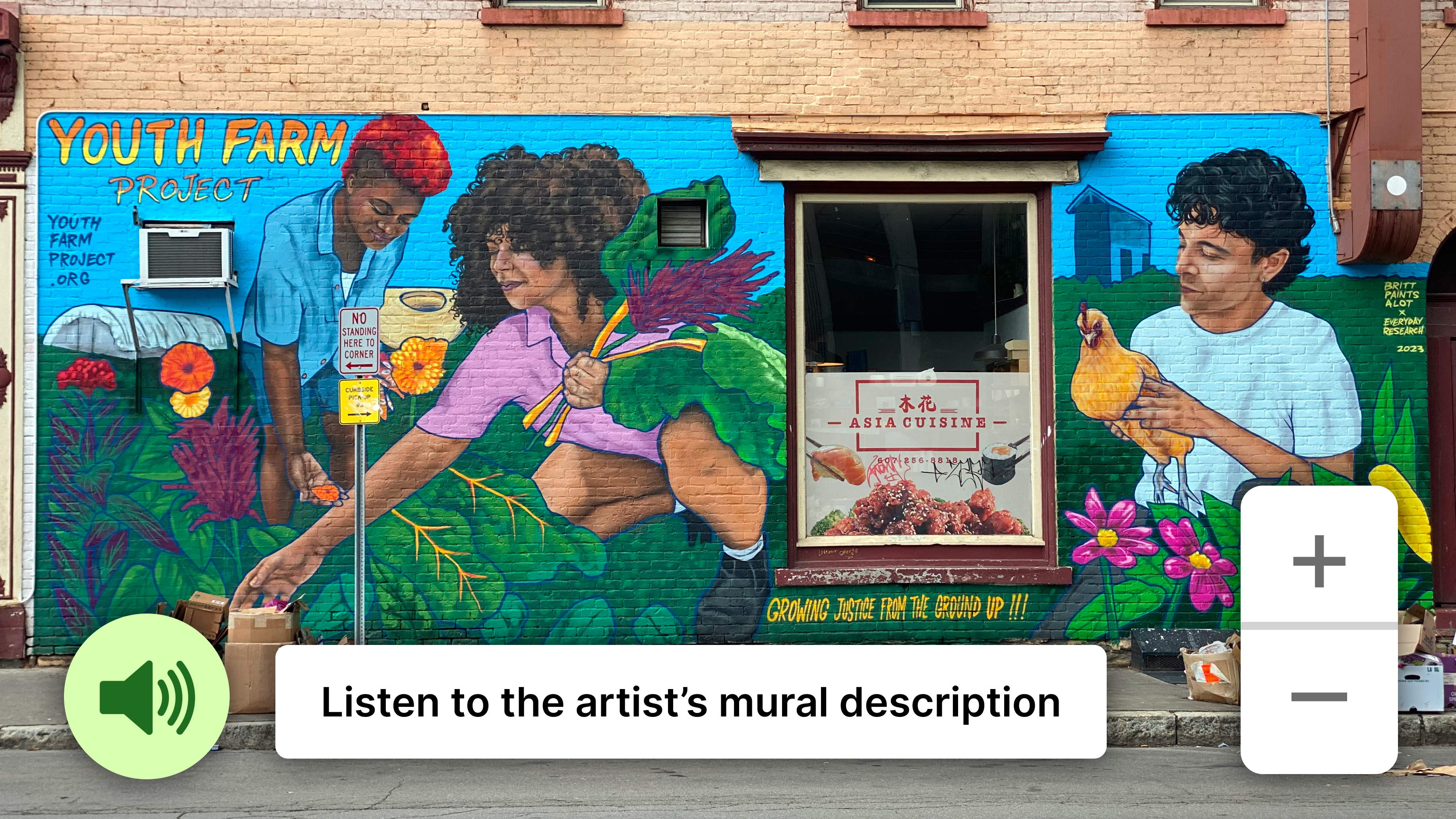}
  \caption{A screenshot of a potential interface for our artifact, showing the Youth Farm Project mural overlaid with three augmented reality components. An early prototype of the visual and auditory experience is presented in this \href{https://www.youtube.com/watch?v=cw5JxsSwDkY}{YouTube video}.}
  \Description{An image of the Youth Farm Project mural, a large mural depicting three young people of color farming. The mural is on the side of a building. At the bottom, from left to right, the controls are a green volume icon, a rectangular bubble saying “Listen to the artist’s mural description,” and a set of plus and minus icons.}
  \label{fig:teaser}
\end{teaserfigure}

\maketitle

\section{Introduction}
As Americans for the Arts states, “public art is a distinguishing part of our public history and our evolving culture. It reflects and reveals our society, adds meaning to our cities and uniqueness to our communities. Public art humanizes the built environment and invigorates public spaces... Public art is freely accessible” \cite{americansforthearts}. Public art can convey a variety of messages \textemdash{} for example, Anish Kapoor’s “Cloud Gate” (colloquially known as “The Bean”) evokes inner reflection \cite{thebean}, while Banksy’s “Love Is In The Air” criticizes militarism and war \cite{sothebys}. These disparate examples also highlight how public art can be created \textit{with} (e.g., a large-scale sculpture funded by private donors) or \textit{without} permission (e.g., graffiti art created by an intentionally anonymous artist). 

Though public art is intended to be accessible to all, in the absence of visual descriptions, it is largely inaccessible to blind and low vision (BLV) people. Some organizations have begun tackling this issue. For example, VocalEye is a non-profit organization based in Vancouver, Canada which is dedicated to providing greater access to theater, arts, and cultural events for BLV people of all ages \cite{vocaleye}. Since 2021, VocalEye has organized multiple crowdsourcing initiatives, dubbed “Crowders,” to encourage the public to submit descriptions for public art events such as the Vancouver Mural Festival. Inspired by my participation in their prior Crowders, in November 2022, I initiated a project to extend VocalEye’s mural accessibility crowdsourcing efforts from Canada to the United States. In collaboration with Ithaca Murals, a non-profit art and activism organization, we engaged both BLV and sighted people in this crowdsourced urban accessibility project. The Crowder yielded 113 descriptions for 14 works of public art in downtown Ithaca, New York, submitted by 25 participants across the U.S. and Canada. In October 2023, we also co-led two in-person mural tours with approximately 25 attendees to share the crowdsourced descriptions with the local community. 

The primary outcomes from our Ithaca Crowder project are (1) a diverse dataset of crowdsourced mural descriptions, (2) two synchronous mural tours, during which we shared mural descriptions with in-person attendees, and (3) increased awareness about art accessibility. However, access to public art should not be a privilege limited to those who could attend our in-person mural tours. We believe that augmented reality (AR) technologies can transform these access measures from ephemeral to enduring. 

As such, we present ARtivism: an AR application for accessible public art and advocacy. This position paper presents a \textit{process artifact}, which consists of an early audiovisual prototype of the experience and a labeled map of specific mural locations. The artifact is informed and inspired by prior work on art, image, and video accessibility (e.g., \cite{li2023understanding, rector2017eyes, cavazos2018interactive, cavazos2021accessible, morris2018rich, stangl2020person, stangl2021going, jiang2022co}) and accessible augmented reality and immersive spaces (e.g., \cite{jiang2023beyond, mott2019accessible, coughlan2017ar4vi, ahmetovic2021musa, guedes2020enhancing, chang2024sound}).

The impact of accessible public art extends beyond making the art itself accessible to blind and low vision people \textemdash{} augmented reality has the potential to improve accessibility to both the aesthetic and advocacy components of public art. In this position paper, we take a first step towards designing an AR-based application for accessible art, grounded in our prior experiences and data from our Ithaca Crowder project. We conclude by exploring the intersections of public art, activism, and emerging technology.

\section{Related Work}
This work is situated at the intersection of two accessibility spaces: (1) art and (2) augmented reality and immersive spaces. To foreground work on art accessibility, we also build on research related to image and video accessibility methods. For example, some have put forth a variety of guidelines and research recommendations for image description experiences in different contexts \cite{stangl2020person, stangl2021going, morris2018rich}, while others have identified what BLV people wish to include in high quality, engaging video descriptions \cite{jiang2022co, natalie2021efficacy, jiang2024s}. 

\subsection{Art Accessibility}
Most prior work on accessible visual art has primarily focused on fine art in a museum setting. Some researchers have investigated accessible designs for museum experiences \cite{luo2023wesee, asakawa2019independent}. Others have focused specifically on understanding how to improve art accessibility through descriptions, music, earcons, and tactile elements \cite{rector2017eyes, cavazos2018interactive, cavazos2021accessible, li2023understanding, kwon2022supporting}. For example, Li and Zhang et al. \cite{li2023understanding} studied BLV people’s experiences with appreciating fine art in museums or art galleries. They found that participants valued having audio descriptions for high level overviews and tactile graphics for fine details, and they also wished to have increased flexibility to seek further details. Kwon et al. \cite{kwon2022supporting} assessed the efficacy of crowdsourcing art descriptions from the general public, as opposed to art experts, finding that this method yielded sufficiently detailed descriptions that allowed BLV people to appreciate and interpret paintings independently. Additionally, Rector et al. \cite{rector2017eyes} leveraged proxemic audio to create interactive sonic experiences with artwork. Through a lab-based study and a live art installation, the authors found that including four layers of detail ranging from background music to detailed verbal descriptions improved BLV people’s overall immersion with the artwork. We draw upon these prior works informing what details and modalities BLV people prefer for art accessibility to explore the nuances of public art and its often physical, site-specific nature.

\subsection{Augmented Reality and Immersive Spaces}
Researchers have also explored how to better design augmented reality systems to better support blind and low vision people. Ahmetovic et al. \cite{ahmetovic2021musa} developed MusA, an AR mobile app to improve visual art accessibility through audio descriptions and visual augmentations of the artwork. BLV art patrons appreciated the zoom and visual augmentation features, and favored the AR application over existing audio guides. In a study with BLV AD users and creators, Jiang et al. \cite{jiang2023beyond} also examined BLV people’s preferences for 360° video accessibility, finding that different senses, such as touch, taste, and smell, could impact their sense of immersion and engagement within a virtual space. Chang et al. \cite{chang2024sound} investigated how BLV people navigated complex sound interactions in mixed reality, given that they include both real-world and virtual-reality audio elements. They proposed six \textit{sound unblending} techniques, including modifying sound characteristics (e.g., pitch, volume), shifting the location of sound sources, and adding earcons. The authors found that these sound manipulations improved sound awareness for BLV users and reduced cognitive load compared to full transparency and noise canceling modes. In this work, we leverage these concepts of visual augmentations, additional sensory outputs, and sound unblending to guide our design for the presentation of description information.

\section{Artifact}
\begin{figure}
  \includegraphics[width=\textwidth]{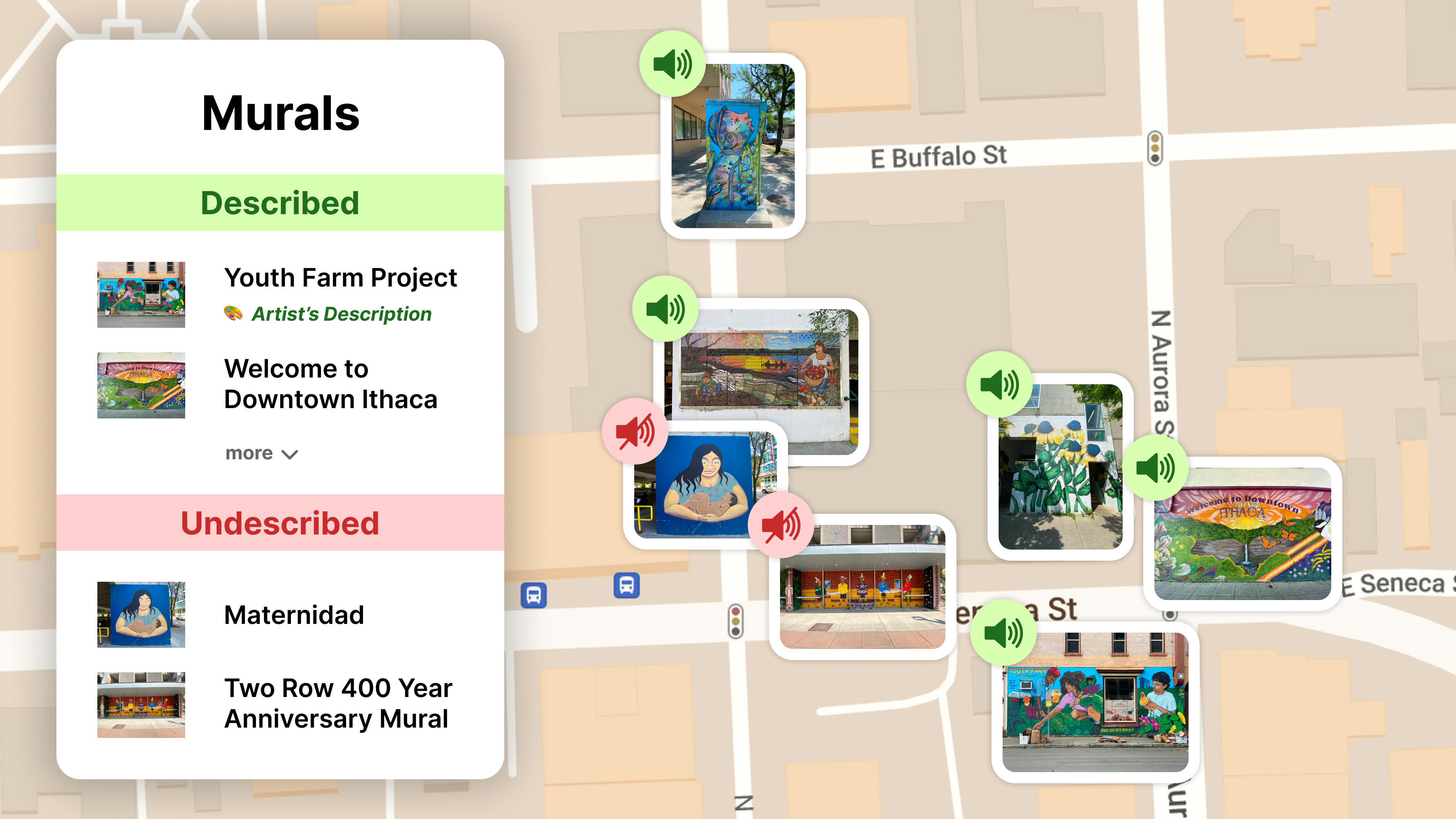}
  \caption{A screenshot of our artifact, showing a map of downtown Ithaca overlaid with images of murals at their specific sites.}
  \Description{A top-down map of Ithaca with a text menu overlaid on the left third of the screen. The menu reads: Murals. Described: Youth Farm Project (Artist’s Description), Welcome to Downtown Ithaca, more. Undescribed: Maternidad, Two Row 400 Year Anniversary Mural. The map has a beige-colored background and streets are delineated in white. There are seven murals represented on this map, five of which have a green volume icon button at the top left of the thumbnail and two of which have a red volume icon button at the top left of the thumbnail.}
  \label{fig:artifact1}
\end{figure}

We designed ARtivism, an AR-based application to make public art and activism more accessible to blind and low vision people. The \textit{process artifact} presented in this position paper consists of early prototypes of a mural experience (Figure \ref{fig:teaser}) and a site map (Figure \ref{fig:artifact1}). We also present a soundscape representing a potential visual and auditory experience (\href{https://www.youtube.com/watch?v=cw5JxsSwDkY}{YouTube link}), demonstrating how the visual information could be conveyed to a BLV user. The application is intended to be flexible and can be used at any location, such as the site of the art or a user’s home. In this section, we provide greater detail about our prototype examples and describe how we came to this design.

\subsection{Prototype}
Figure \ref{fig:teaser} shows a screenshot of a potential visual interface for the artifact, featuring the Youth Farm Project mural, one of the social justice-focused murals from our Ithaca Crowder project. The interface overlays the view of the mural with three augmented reality components on the bottom half, which are a green icon indicating the existence of a description for this mural, a textbox explaining that the description for this mural was written by the artist themself, and a set of plus and minus icons to allow participants to zoom and in out from the mural (as suggested by prior work \cite{ahmetovic2021musa}). 

The prototype of the visual and auditory experience is presented in this \href{https://www.youtube.com/watch?v=cw5JxsSwDkY}{YouTube video}. For our prototype, one of the artists of the Youth Farm Project mural wrote and narrated the description. We recommend involving the artists in writing and recording their descriptions for all public art, if possible. For any works that do not have participant-submitted narrations, we believe it is invaluable to involve BLV people in the narration process to humanize the descriptions, as opposed to using text-to-speech services or other automated systems.

Our site map (Figure \ref{fig:artifact1}) shows a subset of the murals featured in the Ithaca Crowder project. Through a menu and the icons on the map, the interface indicates which murals are and are not described. Descriptions are crowdsourced from the public to capture a wide variety of perspectives on the artwork. Though respondents are not necessarily art experts, the submission form encourages them to follow art and image accessibility best practices and includes some open-ended prompts to elicit both objective and subjective reactions. To ensure accessibility for BLV users, the thumbnail photos on the map will have alt-text containing the artwork’s name, a brief, high-level description of the art (regardless of whether it has a longer description), and the mural’s approximate location. 

\subsection{Design Rationale}
\subsubsection{Why head-mounted augmented reality?}
Researchers have already begun exploring how augmented reality can improve accessibility for BLV people (e.g., \cite{coughlan2017ar4vi, chang2024sound, ahmetovic2021musa}). AR is not limited to head-mounted displays \textemdash{} mobile phones are also commonly used due to their ubiquity. However, as found in prior work on mobile-based AR and art exhibits, holding up a phone for the entire duration of a lengthy description can be cumbersome \cite{ahmetovic2021musa}. Therefore, for this design, we chose to use a similar form factor to prior AR technologies for accessibility, such as wearable OrCam devices \cite{orcam}. Especially as head-mounted displays become more and more common, we speculate that this may normalize the use of such technologies among both BLV and sighted patrons for accessing more information about artwork.

\subsubsection{What are the benefits of crowdsourcing?}
Kwon et al. \cite{kwon2022supporting} demonstrated the viability of crowdsourcing art descriptions from non-experts in a museum setting. For public art with an activism focus, crowdsourcing may be even more critical. A description that might capture the visual appearance of a work of art may not highlight or even surface the actual underlying message of the art. Furthermore, different people may have different perspectives or reactions to an artwork. We believe that compiling a diverse set of perspectives about a single work of art can enhance BLV people’s experiences with independently appreciating art. However, we also encourage describers to avoid prescribing or overexplaining the meaning behind any artwork.

\subsubsection{How will the output be designed?}
As with prior work, we posit that AR goes beyond visual changes \textemdash{} AR can also include auditory, tactile, olfactory, and gustatory modifications to the physical world \cite{jiang2023beyond}. When viewing public art, it can be valuable to hear the surrounding soundscape to better contextualize why a work of art may be in a certain location (e.g., why a commemorative sculpture of a local woman is located within the heart of Ithaca’s downtown area). We consider how concepts of sound unblending \cite{chang2024sound} can be helpful for reducing cognitive load while also preserving important contextual information about where a mural is situated within an urban setting. We also recommend for future work to leverage advanced AI models to create the requisite files to support tactile output from refreshable Braille displays \cite{jiang2024s}. Ultimately, the design of the output should be left for a user to fine-tune, but prior guidelines can provide a starting point for designers, creators, and artists when considering how to craft these accessible and engaging art experiences.

\section{Discussion}
The act of creating an application to store descriptions and public art locations inherently surfaces what art is where. However, we must consider that not all art is “permitted” \textemdash{} while a city can commission a mural on a particular building, graffiti artists may not want their activity to be logged on an application out of privacy and enforcement concerns. Banksy is one prominent example of an anonymous graffiti artist whose works often focus on social justice. In such cases, we encounter a difficult tension between providing access to art and advocacy for BLV audiences while also respecting the wishes of the artist. Although we do not yet have an answer to this question, we believe that this is an important consideration and distinction that is specific to public art and other forms of artwork that are not located within a formal gallery setting (i.e., “without permission”). We encourage future work to consider solutions to make unsanctioned artwork more accessible without compromising artists’ privacy or safety.

\section{Conclusion}
In this position paper, we presented a \textit{process artifact} showcasing an AR-based application for making art and advocacy more accessible to blind and low vision people. We situate our work at the intersection of art and augmented reality accessibility, and explore the nuances associated with the site-specific and often grassroots nature of public art. We hope that this paper provides a preliminary step towards exploring the potential for AR to make public art and creative visual expression more accessible to blind and low vision people, thereby increasing access to activism efforts as well. 

\begin{acks}
We thank \href{https://www.vocaleye.ca/}{VocalEye}, a Canada-based non-profit organization dedicated to making the arts more accessible to blind and low vision people; Caleb R. Thomas, the founder and organizer of \href{https://www.ithacamurals.com/}{Ithaca Murals}; and Valerie Foster Githinji, an Advocacy and Care Specialist at the \href{https://fliconline.org/FLIC/}{Finger Lakes Independence Center}. Without them, this project would not have been possible. We also thank Daniel Zhu, Kelly Jiang, and Sophia Hwang for their feedback on this position paper. Lastly, we are grateful to \href{https://www.everydayresearch.com/}{Efren Rebugio, Jr.} and \href{https://brittpaintsalot.com/}{Brittany Johnson} for painting the Youth Farm Project mural, writing the description, and providing the narration for the prototype video. 
\end{acks}

\bibliographystyle{ACM-Reference-Format}
\bibliography{references}

\end{document}